\documentclass[nofootinbib]{article}
\usepackage[total={6.5in,9in},top=1in,headsep=0.1in,headheight=1in]{geometry}
\usepackage[utf8]{inputenc}
\usepackage{hyperref}
\usepackage{amssymb}
\usepackage{graphicx}
\usepackage{lipsum}
\usepackage{hyperref}
\usepackage{placeins}
\usepackage[maxbibnames=10,style=numeric-comp,sorting=none]{biblatex}
\usepackage{upgreek}
\hypersetup{hidelinks}
\usepackage[colorinlistoftodos]{todonotes}
\usepackage[hybrid]{markdown}
\usepackage{color}
\usepackage{enumitem}
\usepackage[caption=false]{subfig}
\usepackage{lineno}
\usepackage[caption=false]{subfig}
\usepackage{graphicx}
\usepackage{wasysym}

\newcommand\snowmass{
\begin{center}
  \rule[-0.2in]{\hsize}{0.01in}\\
  \rule{\hsize}{0.01in}\\
  \vskip 0.1in
  Submitted to the Proceedings of the US Community Study\\ 
  on the Future of Particle Physics (Snowmass 2021)\\
  \rule{\hsize}{0.01in}\\
  \rule[+0.2in]{\hsize}{0.01in}\\[-2em]
\end{center}
}

\usepackage[firstpage=true]{background}
\backgroundsetup{contents={\parbox{6.5in}{\snowmass}}, scale=1,placement=top,opacity=1,color=black,position={3.25in,1.2in}}

\usepackage{fancyhdr}
\fancypagestyle{plain}{%
  \fancyhf{}%
  \fancyhead[C]{}
  \fancyfoot[C]{\thepage}
}

\fancypagestyle{empty}{%
  \fancyhf{}%
  \fancyhead[C]{{\it Snowmass 2021 Dark Matter Complementarity Report}}
  \fancyfoot[C]{\thepage}
}
\title{Dark Matter Complementarity - Summary}
\date{\today}

\usepackage{authblk}

\author[1]{Antonio Boveia}

\author[2]{Mohamed Berkat}

\author[3]{Thomas Y. Chen}

\author[ \hspace{-1ex}]{Aman Desai}

\author[2,4]{Caterina Doglioni}

\author[5,6,7]{Alex Drlica-Wagner}

\author[8]{Susan Gardner}

\author[9]{Stefania Gori}

\author[2]{Joshua Greaves}

\author[10]{Patrick Harding}

\author[11]{Philip C. Harris}

\author[12]{W. Hugh Lippincott}

\author[13,14,15]{Maria Elena Monzani}

\author[16]{Katherine Pachal}

\author[17]{Chanda Prescod-Weinstein}

\author[18]{Gray Rybka}

\author[19]{Bibhushan Shakya}

\author[20]{Jessie Shelton}

\author[21]{Tracy R. Slatyer}

\author[22]{Amanda Steinhebel}

\author[23]{Philip Tanedo}

\author[13]{Natalia Toro}

\author[13]{Yun-Tse Tsai}

\author[11]{Mike Williams}

\author[11]{Lindley Winslow}

\author[24]{Jaehoon Yu}

\author[25]{Tien-Tien Yu}

\affil[1]{Department of Physics and Center for Cosmology and Astroparticle Physics, The Ohio State University,
191 W. Woodruff Avenue Columbus, OH 43210, USA}

\affil[2]{Fysiska institutionen, Lunds universitet, Professorsgatan 1, Lund, Sweden}

\affil[3]{Fu Foundation School of Engineering and Applied Science, Columbia University, New York, NY 10027, USA}

\affil[4]{University of Manchester, Department of Physics and Astronomy, Manchester M13 9PL, United Kingdom}

\affil[5]{Fermi National Accelerator Laboratory, Batavia, IL 60510, USA}
\affil[6]{Kavli Institute for Cosmological Physics, University of Chicago, Chicago, IL 60637, USA}
\affil[7]{Department of Astronomy and Astrophysics, University of Chicago, Chicago IL 60637, USA}

\affil[8]{Department of Physics and Astronomy, University of Kentucky, Lexington, KY 40506-0055}

\affil[9]{Physics Department, University of California, Santa Cruz, CA 95064, USA}

\affil[10]{Physics Division, Los Alamos National Laboratory, Los Alamos, NM 87545, USA}

\affil[11]{Laboratory for Nuclear Science, Massachusetts Institute of Technology, Cambridge, MA 02139, USA}

\affil[12]{Department of Physics, University of California, Santa Barbara, CA 93106, USA}

\affil[13]{SLAC National Accelerator Laboratory, Menlo Park, CA 94025, USA}

\affil[14]{Kavli Institute for Particle Astrophysics and Cosmology, Stanford University, Stanford CA, USA}

\affil[15]{Vatican Observatory, Castel Gandolfo, V-00120, Vatican City State}

\affil[16]{TRIUMF, Vancouver, BC V6T 2A3, Canada}

\affil[17]{Department of Physics and Astronomy, University of New Hampshire, Durham, NH 03824, USA}

\affil[18]{Department of Physics, University of Washington, Seattle WA 98195, USA}

\affil[19]{Deutsches Elektronen-Synchrotron DESY, Notkestr.~85, 22607 Hamburg, Germany}

\affil[20]{Department of Physics, University of Illinois Urbana-Champaign, Urbana, IL 61801}

\affil[21]{Center for Theoretical Physics, Massachusetts Institute of Technology, Cambridge, MA 02139, USA}

\affil[22]{NASA Goddard Space Flight Center, Greenbelt, MD, USA}

\affil[23]{Department of Physics and Astronomy, University of California Riverside, Riverside, CA 92521, USA}

\affil[24]{Physics Department, University of Texas, Arlington, TX 76019, USA}

\affil[25]{Department of Physics and Institute for Fundamental Science, University of Oregon, Eugene, OR 97403, USA}

\begin{document}
\maketitle

\begin{abstract}
The fundamental nature of Dark Matter is a central theme of the Snowmass 2021 process, extending across all Frontiers. In the last decade, advances in detector technology, analysis techniques and theoretical modeling have enabled a new generation of experiments and searches while broadening the types of candidates we can pursue. Over the next decade, there is great potential for discoveries that would transform our understanding of dark matter. In the following, we outline a road map for discovery developed in collaboration among the Frontiers. A strong portfolio of experiments that delves deep, searches wide, and harnesses the complementarity between techniques is key to tackling this complicated problem, requiring expertise, results, and planning from all Frontiers of the Snowmass 2021 process. 
\end{abstract}

\newpage

\section*{Executive Summary}
The evidence for Dark Matter (DM) is overwhelming, yet the fundamental nature of its constituents remains a mystery. Over the last decade, we have built a powerful and diverse collection of tools to unlock this mystery, both by refining established technologies and techniques and by harnessing new ones including artificial intelligence/machine learning (AI/ML) and quantum sensing/control. In parallel, we have continued to build our understanding of how DM shapes our universe. We are well-positioned for a great discovery.

From its production to its interactions, DM is a major science driver across all experimental Frontiers -- Cosmic Frontier (CF), Energy Frontier (EF), Neutrino Frontier (NF), and Rare Processes and Precision Frontier (RF) -- as well as the cross-cutting Frontiers: Accelerator Frontier (AF), Community Engagement Frontier (CEF), Computational Frontier (CompF), Instrumentation Frontier (IF), Underground Facilities (UF) and Theory Frontier (TF). Because the science of DM does not respect Frontier boundaries, a unified strategy is needed to maximize discovery potential.

\noindent
{\bf Complementarity Within and Across Frontiers}

Complementarity drives discovery in multiple ways. The space of viable DM candidates and their properties is large, and  a single experimental technique or approach cannot be used to test all the possibilities; a diverse range of techniques provides access to a much broader ensemble of DM scenarios. Where different approaches have simultaneous sensitivity to a particular DM candidate, they provide essential and complementary information and promote healthy competition. 

Some techniques can tell us whether a new particle constitutes the bulk of the DM in the Galactic halo while others may better elucidate interactions of DM with known particles, with a third category mapping the spectrum of new ``dark sector'' particles related to the DM. 
In the event of a discovery, detection and exclusion by complementary techniques will help triangulate the fundamental nature of DM.

\noindent
{\bf Maximize Opportunities for Discovery: Delve Deep, Search Wide}

Embracing the role of complementarity, the DM community proposes a strategy to \emph{delve deep and search wide} to maximize discovery potential. A range of highly compelling theoretical targets arising from simple/minimal models are accessible in the next decade via planned and proposed CF, EF, NF, and RF experiments, colliders, and observatories. 
While discovering the fundamental nature of DM is the ultimate prize, searching in these regions and \emph{not} finding DM would provide important information on the properties of DM. 
Simultaneously, our strategy encompasses the development of new technologies and techniques to explore new possibilities and complement the sensitivity of existing searches.

\noindent
{\bf Discovery Strategy}\\
The community puts forth the following strategy for discovering the fundamental nature of DM:

\noindent $\newmoon$~ \textbf{Build a portfolio of experiments of different scales:}
Experiments at all scales are needed to untangle the mystery of DM and cover the very broad range of theoretically motivated parameter space. Existing and planned large-scale facilities across the HEP Frontiers have exceptional potential to discover fundamental properties of DM.
We should commit to scaling up mature technologies that can promise significant sensitivity improvements, developing potentially transformative new technologies to maturity, and supporting efforts to maximize and make accessible large projects' science output in the search for DM.
At smaller scales, execution of the existing Dark Matter New Initiatives (DMNI) program and similar future calls are necessary to build the most compelling DM portfolio, develop experience in project execution, and accelerate the pace of discovery.

\noindent $\newmoon$~ \textbf{Leverage US expertise in international projects:}
The effort to understand the fundamental nature of DM is a world-wide endeavor. Coordination and cooperation across borders is critical for enabling this discovery. While building a strong US-based program, we should pursue opportunities to leverage key US expertise as a collaborative partner in international projects and play a leadership role in this critical area.  

\noindent $\newmoon$~ \textbf{Provide support to further strengthen the theory program:}
A strong theory program is essential to make connections between experimental Frontiers and take full advantage of new developments in analysis techniques. Theorists' input has been and will be critical for developing innovative new approaches to better understand and detect DM, and for determining how to predict and relate signals across a range of experimental probes.

\noindent $\newmoon$~ \textbf{Support inter-disciplinary collaborations that enable discovery:}
Many searches for DM benefit greatly from cross-disciplinary expertise, with examples ranging from nuclear physics to metrology, and astrophysics to condensed matter and atomic physics. Mechanisms to support such inter-disciplinary collaborations should be established.

\noindent $\newmoon$~ \textbf{Targeted increase in the research budget:} 
New research funding targeted toward solving the DM problem is essential to enable new ideas, new technologies, and new analyses. The number of active efforts exploring DM has increased tremendously in the past decade, without a concomitant increase in research funding. 
Across all Frontiers and project scales, research funding is critical to enable discovery and build on new capabilities, both in projects focused specifically on DM and to support DM analyses at multi-purpose experiments.
Without such support, the community will not be able to execute the program described here, decreasing the chances of solving the mystery of DM.



\section{Introduction}

Determining the fundamental nature of dark matter (DM) is one of the major open questions that confronts our understanding of physics, and it has been among the guiding themes of the Snowmass process in most HEP Frontiers. While the microscopic properties of DM remain almost completely unknown, the relatively similar energy densities of dark and visible matter in the Universe --- DM has five times more energy density than visible matter --- is suggestive that there may be non-gravitational interactions between DM and the Standard Model (SM). However, the nature of these interactions is essentially unconstrained, necessitating a broad and comprehensive approach to this question to make progress in the next decade. 
This challenge requires expertise, results, and planning from the full range of communities involved in the Snowmass 2021 process, across all Frontiers.

The 2013 Snowmass process had a topical group (CF4) specifically devoted to the complementarity of different DM studies. 
The white paper produced by that group, \textit{``Dark Matter in the Coming Decade: Complementary Paths to Discovery and Beyond,"}~\cite{Snowmass2013CosmicFrontierWorkingGroups1-4:2013wfj}
reviewed existing and planned DM efforts in direct detection, indirect detection and collider experiments, as well as in astrophysical probes. 

The need for diverse and complementary approaches to the DM problem is even more pressing now than it was in 2013.
The DM search domain has broadened significantly with promising new avenues now under development that will yield results in the next decade.
Our theoretical understanding of possible DM parameter space has grown, and that space is being explored by a much larger number of projects at different scales. Working in tandem, experiment, observation, theory, and computation have the potential to identify key DM properties while definitively excluding vast swathes of parameter space.

This document
re-casts the scope and definition of complementary approaches to DM identification to reflect the current state of the field (Section \ref{sec:DMComplementarity}), 
 summarizes the needs of the different communities looking for DM and their complementary strengths (Section \ref{sec:IndividualTGComplementarityNeeds}),
and supports  these arguments with cross-Frontier case studies (Section \ref{sec:CaseStudies}). An extended version of this report is also available \cite{Boveia:2022syt}.

\section{Dark matter complementarity in the coming decade}
\label{sec:DMComplementarity}

A complementary ensemble of DM searches is key to a comprehensive strategy, with advantages including:

\textbf{1. Different approaches to DM searches allow us to probe different fundamental properties of the DM.} For this reason, complementary approaches will be necessary to fully identify the DM and understand its physics. For example, cosmological and astrophysical probes allow access to environments not found on Earth and time/space scales dwarfing terrestrial experiments, and consequently have unique sensitivity to a range of properties including the DM lifetime, annihilation rate, and self-interaction cross section. Rare interactions between DM and the SM are often most precisely probed in terrestrial experiments, which enjoy controllable and clean environments. DM production could be observed in extreme environments in the Cosmic Frontier, or under controlled laboratory conditions at accelerators and colliders in the Energy, Rare Processes and Precision, and Neutrino Frontiers; such measurements can reveal DM interactions in energy domains beyond those of the halo DM, including its early-universe behaviour and any dark sector particle spectrum beyond DM.  This kind of complementarity is showcased in the case studies \textit{Minimal WIMP Dark Matter}, involving the Energy and Cosmic Frontiers, \textit{Sterile Neutrino Dark Matter}, involving the Cosmic and Neutrino Frontiers, and \textit{Wave-like Dark Matter: QCD Axion Discovery}, involving the Cosmic, Neutrino, and Rare Processes Frontiers. 

\textbf{2. Different approaches to DM searches offer unique discovery sensitivity to distinct scenarios and regions of parameter space.} Only by performing a variety of experiments, covering multiple Frontiers, can we span the wide DM parameter space. For example, DM interactions with the SM may be either suppressed or enhanced at low energies; to maximize the chances of a discovery, it is important to support both experiments sensitive to non-relativistic DM signals and those that explore the physics of DM at higher (often relativistic) energies. This kind of complementarity is also explored in the case study \textit{Generic BSM-mediated and Vector Portal Dark Matter}, involving the Energy and Rare Processes and Precision Frontiers.

\textbf{3. Results from any one class of searches can continuously inform the interpretation of other measurements.} A Cosmic Frontier detection of relic DM would connect any observed candidate to the cosmological evidence that motivates the entire DM program, and provide a target and motivation for future efforts in other Frontiers. Combining results obtained with different approaches can address otherwise-intractable uncertainties --- e.g.~DM production probes are independent of the fraction of DM due to a particular candidate, or the DM distribution. Complementary measurements can also reduce systematic uncertainties directly: as two examples, cosmic measurements of the density and velocity distribution of DM are essential for the interpretation of direct and indirect searches, and accelerator experiments can constrain cosmic ray physics relevant for indirect searches~\cite{Cooley:2022ufh}. 

\textbf{4. Different DM experiments can be co-located and/or profit from the same or similar technological infrastructure.} Efficient use of shared resources enables a wider exploration of DM. Examples include small Rare Processes and Precision Frontier accelerator experiments that can be co-located with Energy Frontier collider experiments, using the same high-energy beams to produce different kinds of DM. Such experiments are discussed in the Rare Processes and Precision Frontier and their connection to collider experiments is discussed in the Energy Frontier report~\cite{Energy-Frontier-Report}.

\vskip 1mm

\textbf{The community searching for DM has grown much more diversified in terms of technologies, search targets, and project scales.} 
Since the last iteration of Snowmass in 2013, many new approaches to searching for DM, as well as new theoretical hypotheses, have attained sufficient maturity to be part of the toolkit that we will use to make progress in the quest for DM in the next decade. 
The older approach to DM complementarity (as outlined in the previous Snowmass whitepaper \cite{Snowmass2013CosmicFrontierWorkingGroups1-4:2013wfj}) focused primarily on the Weakly Interacting Massive Particle (WIMP) hypothesis and direct DM-SM interactions, and incorporated high-mass new particle searches, direct and indirect detection, and astrophysical probes. This remains an important hypothesis that should be rigorously tested as part of a program that ``delves deep.'' However, the WIMP is now joined by a greater diversity of alternative DM candidates. The QCD axion has emerged as another key target for a focused suite of experiments that will enable a definitive search for this candidate. Candidates and possible signatures that inform the wider strategy include light particle-like DM with masses in the MeV-GeV range, wave-like DM in addition to the QCD axion (such as ultralight scalar or vector bosons), signatures of the greater dark sector including long-lived particles, cosmological observations of DM properties on large scales, and new signatures of DM in gravitational waves and other multi-messenger observations.

The strategy presented above was developed from the bottom-up through the communities represented by the Frontier Topical Groups. It is well-aligned with the Basic Research Needs for Dark Matter Small Projects New Initiatives report \cite{BRNreport}, which highlighted the growing landscape of smaller experiments to produce and detect DM at accelerators, as well as the direct detection of light and ultra-light DM. The goal of the following sections is to summarize the needs of the individual Topical Groups and show that the needs are highly complementary as are the techniques. A common strategy across HEP is needed to enable the discoveries that will reveal the nature of DM.

\section{Realizing dark matter complementarity across Frontiers}
\label{sec:IndividualTGComplementarityNeeds}

In this section, we briefly summarize the main approaches towards identifying the fundamental nature of DM from each Snowmass Frontier,
referring to the Topical Group whitepapers for further information and needs.

\paragraph{Cosmic Frontier 1 - Direct and Indirect Searches for Particle Dark Matter.}

\noindent
Direct detection experiments using mature technologies, like large liquid noble detectors, probe some of the smallest cross sections ever measured in the non-relativistic regime --- favorable for models where interactions are enhanced at low velocities --- and provide leading sensitivity to heavy DM up to ultraheavy mass scales.  A new era of technological development has enabled sensitivity to tiny energy depositions of eV-scale and below, opening up sensitivity to unexplored parameter space for DM masses below 1 GeV.  
Indirect detection provides a model-independent probe of the minimal thermal relic scenario with $s$-wave annihilation, with expected sensitivity up to tens of TeV masses in the next decade, and constraints of some form up to the Planck mass. Such searches provide unique probes of the DM decay lifetime (and other long-timescale processes), via their access to distance and time scales that dwarf any terrestrial experiment, and cover enormous mass ranges.

The next decade offers a broad array of exciting opportunities in direct and indirect detection, as discussed in Ref.~\cite{Cooley:2022ufh}.
To pursue these opportunities requires a diverse, continuous portfolio of experiments that includes both direct and indirect detection techniques at multiple scales. Moderate- and large-scale experiments allow us to delve deep into high priority target scenarios such as WIMPs, whereas an ensemble of small-scale experiments like those supported by the Dark Matter New Initiatives (DMNI) program provides versatility and the ability to test an expanded range of models. Support for theory, simulations, calibration, background modeling and complementary astrophysical measurements is essential to enabling discovery, as is R\&D towards improved detector technologies. Lastly, direct detection experiments, particularly the next generation of WIMP searches, will require continued investment in underground facilities.

\paragraph{Cosmic Frontier 2 - Direct Searches for Axion Dark Matter.}
\noindent
Wave-like DM encompasses all candidates with masses less than 1\,eV. Due to their small masses, the detection principles are vastly different than those traditionally used in high energy physics. It is here where quantum measurement techniques become critical and advancements in this area have opened up a broad horizon of new candidates to explore and many opportunities for discovery. 
Within this group, the well-motivated QCD axion is an excellent DM candidate that also solves the strong CP problem. Building on the success of ADMX-G2, the new moderate-scale DMNI experiments ADMX-EFR and DMRadio-m$^3$ are readying to start construction. In the next decade, a portfolio of moderate-scale experiments are poised to explore significant QCD axion parameter space. A concentrated effort coupling R\&D, demonstrator-scale experiments, and theory, would enable searches for a broader spectrum of candidates.

\paragraph{Cosmic Frontier 3 - Cosmic Probes of Dark Matter.}
\noindent
Cosmic probes provide the only direct, positive empirical measurements of the existence and properties of DM.
These probes complement terrestrial DM searches by constraining the interaction strength between DM and the SM in otherwise inaccessible regions of parameter space.
In addition, cosmic probes provide the only known way to directly study the fundamental properties of DM through gravity, the only force to which DM is known to couple.
Cosmic probes are sensitive to the DM mass, lifetime, self-interaction cross section, and other dark sector particles.
In particular, cosmic probes are on the cusp of detecting DM halos that are devoid of baryonic galaxies, providing a strong test of the cold, collisionless DM paradigm.

The construction of future facilities spanning the electromagnetic spectrum, as well as gravitational waves, can provide sensitivity to DM physics, as well as the physics of dark energy and the early universe. 
Strategic HEP investments in the construction and operation of Rubin LSST, CMB-S4, and Spec-S5 should include DM physics as a core science driver to be considered during the design and operation of these experiments.
Cosmic probes provide robust sensitivity to the microphysical properties of DM due to enormous progress in theoretical modeling, numerical simulations, and astrophysical observations. 
Theory, simulation, observation, and experiment must be supported together to maximize the efficacy of cosmic probes of DM physics.

\paragraph{Rare \& Precision Frontier 6 - Dark Sectors at High Intensities.}
Intensity Frontier experiments offer unique access to the physics of low-mass DM by systematically probing a broad range of simple, well-motivated dark sectors neutral under SM forces. 
Near-term searches for \emph{DM production} are needed to thoroughly explore the coupling ranges motivated by MeV-to-GeV thermal DM. Because the energies probed in Rare \& Precision Frontier experiments are similar to those relevant for light DM thermal freeze-out, the range of production cross-sections expected for low-mass thermal relics is compact---and accessible---regardless of the DM spin.  In DM models where interactions are suppressed at low velocities, cosmological and Galactic DM signals can be suppressed by orders of magnitude. Thus, accelerator-based production is the favored path to detection of these scenarios. 
Intensity Frontier experiments can also discover and characterize \emph{light dark-sector particles that decay into SM particles}, often with detectably long lifetimes. Such particles arise in most generalized freeze-out scenarios such as strongly interacting massive particles, forbidden DM, and secluded DM~\cite{Asadi:2022njl}. Their discovery can shed light on the interactions, nature, and origin of DM. Both goals, searching for DM production and for visibly decaying particles related to DM, were called out in the 2018 DM New Initiatives Basic Research Needs report~\cite{BRNreport}. 

Achieving these goals over the next decade requires a four-pronged approach: performing dark-sector analyses at multi-purpose experiments; realizing the experiments selected through the competitive DMNI program; broadening the DMNI experimental portfolio to achieve the goals laid out in the DMNI Report, including a focus on signatures of long-lived dark sector particles decaying  (semi)visibly; and continued investment in dark-sector theory.

\paragraph{Energy Frontier 10 - Dark Matter at Colliders.}
Present and future colliders~\cite{Bose:2022obr,Energy-Frontier-Report} can search for particle DM and its interactions, covering a broad swath of scenarios ranging from the canonical WIMP to more general models of DM particles and the mediators of their interaction or extended dark sectors.
They can detect invisible particles produced in collisions via missing transverse momentum, but their greatest strengths are their ability to study how these invisible particles interact with other particles and to search for additional particles involved in the DM physics.

Over the next decade, the High-Luminosity LHC (HL-LHC) can explore whether DM couples to the Higgs boson or other beyond the Standard Model (BSM) portal particles, and test supersymmetric and other particle DM candidates that have been long-term targets of the field.
On longer timescales, an electron-positron collider can push this sensitivity significantly further, as well test DM models that favor couplings to leptons.
A hadron or muon collider would allow direct exploration of far-higher energy scales, 
with the greatest prospects to discover WIMP multiplets, and kinetically-mixed thermal DM above a GeV.

High-energy colliding beam facilities can also be used for special-purpose, co-located DM experiments searching for long-lived, dark sector particles. An example is the case of the proposed Forward Physics Facility \cite{Feng:2022inv} for the HL-LHC, but similar facilities would provide similar capabilities at other future colliders. 

\paragraph{Neutrino Frontier 3 - Dark Matter in Neutrino Experiments.}
The high power beams, multi-kiloton scale far detectors underground, and near detector systems needed for precision neutrino experiments make a broad variety of BSM searches possible in the Neutrino Frontier.
The high power proton beams can produce dark sector particles, such as axion-like particles, light DM and heavy neutral leptons, which can be detected in the precision near detector complex.
Dark sector particles with oscillatory behavior, e.g. sterile neutrinos, can be probed in the far detector in combination with the near detector complex in long baseline neutrino experiments.
Dark sector particles that alter neutrino fluxes from natural sources, or those with cosmogenic origins such as boosted DM, can be detected in the underground, massive far detectors.

Such rare dark sector interactions can be easily masked by backgrounds from a variety of sources, such as cosmic rays, terrestrial radioactive sources, and neutrino interactions.
The current level of understanding of neutrino-nuclear interactions as well as neutrino spectra from natural sources is insufficient to effectively estimate or control them to the precision necessary for longer term dark sector particle searches, necessitating a strong synergistic and sustained collaboration between the nuclear physics and high energy physics communities to perform measurements needed to improve the modeling of neutrino-nuclear interactions, such as the \href{https://www.e4nu.com}{$e4$nu} collaboration. 

\paragraph{Theory Frontier.}
The allowed mass ranges and interactions for DM are so vast that it is impossible to undertake any meaningful exploration of this parameter space without insight from theory. Theory is the language needed to make sense of results from different experiments and Frontiers \cite{Green:2022hhj}. Broadly, theory complements and enhances experimental DM searches in three ways: 

Theory is essential to \emph{define connections between experimental programs and understand their complementarity} in the context of specific DM models. This role broadens the impact of null results, and becomes crucial in understanding and verifying a potential signal. 

Theory can \textit{motivate experimental programs} for DM searches by tying DM to other deep questions and insights about particle physics.  
Examples include the hierarchy problem (a central problem of the Energy Frontier) inspiring WIMP DM, neutrino mass generation (a central problem of the Neutrino Frontier) inspiring sterile neutrino and neutrino-portal DM, and the strong CP problem giving birth to axion DM.

Theory can also \emph{enable new paths to DM detection}, identifying DM scenarios accessible to ongoing experiments or even leveraging new technologies, experimental capabilities, and insights to define new experimental directions. Recent examples of such interplay are found throughout the DMNI program across Rare Processes and Cosmic Frontiers.

\paragraph{Instrumentation Frontier.}
Advances in instrumentation support every aspect of the hunt for DM, with new technologies often opening new regions of parameter space for exploration. Two examples in direct detection are in quantum sensors and noble element detectors. Rapid progress in quantum sensors over the past decade has been key in the success of the wave-like DM program. Such sensors come in a wide range of technologies: atom interferometry, magnetometers, calorimeters, and superconducting sensors to name a few. In searches for WIMP-like DM, the development of liquid noble detectors has paved the way for the huge strides in sensitivity seen since the last Snowmass, with LXe detectors leading the way and LAr close behind. Of course, there are many more examples that enable DM searches in all the Frontiers, including development in photon sensing, timing, calorimetry, etc. Continued R\&D into instrumentation, development of the technical workforce, and tools to share common knowledge will be an important component of the future DM program. More information can be found in the Instrumentation Frontier report~\cite{IFReport}. 

\paragraph{Computational Frontier.}

Computing and software are critical components of any search for DM, from the collection and storage of raw data, through various stages of data movement, data processing and data analysis, all the way to the interpretation of results. Interestingly enough, the majority of DM searches are rapidly converging towards exascale datasets that are not matched by technological developments and budgets~\cite{SmallExperiments,CF1-WG4-Report, CompF-Report,stewart_graeme_andrew_2020_4009114}. Given this dramatic escalation in data volume and complexity, common computing needs can be identified to enable successful DM searches across all Frontiers, including: stronger partnerships with the national supercomputing facilities, lowering the barrier of entry and providing input on architecture evolution; scalable software infrastructure tools across HEP, avoiding duplication of effort; simulation tools that are efficient, well-maintained, well-understood, and thoroughly validated; and industry collaborations on machine learning techniques, leveraging access to external experts.

An additional challenge specific to complementarity is the need to exchange data between different experiments and even Frontiers, which implies the necessity to converge on data formats and analysis tools. The DM community should take the lead in advocating for widely-adopted data and software standards, in support of global analyses of experimental results. Complementarity studies would be strengthened by an Open Science paradigm including both data and software, with the added benefit of enhancing the credibility of our potential discoveries, since progress in the field of DM will require thorough scrutiny of data and results.

\paragraph{Underground Facilities.}
Most DM direct detection experiments must be sited in underground facilities to evade cosmic ray backgrounds, and multiple new underground DM experiments are expected and being planned (at both large and small scales). Currently, underground facilities are largely subscribed by existing projects, with only limited space available in the coming years. There is, then, a clear need for additional underground space, tailored to the needs of DM experiments.
This underground space must accommodate experiments across scales, including large liquid noble or freon experiments and smaller installations, for example mK facilities. Assembly of future experiments will occur largely in the underground environment, requiring underground radon-free clean rooms. Given the volume of gas/cryogen, future liquid noble experiments also require underground areas for staging (e.g., gas storage) and utilities (e.g. pumps, distillation). These new suitable spaces must be available by the late 2020s to meet the demand, which may be met in North America by proposed new excavations at SURF or SNOLAB. More information can be found in the UF report~\cite{Underground-Frontier-Report}.

\paragraph{Accelerator Frontier.}
Current and future accelerator facilities are the underpinning of both Intensity Frontier and Energy Frontier searches for DM and dark sectors. 
Several beam facilities for axion and DM searches have been shown to have great potential for construction in the 2030s in terms of scientific output, cost and timeline, including PAR (a 1 GeV, 100 kW PIP-II Accumulator Ring). In general, we should efficiently utilize existing and upcoming facilities to explore dedicated or parasitic opportunities for rare process measurements --- examples include the SLAC SRF electron linac, MWs of proton beam power potentially available after construction of the PIP-II SRF linac, spigots of the future multi-MW FNAL complex upgrade, and at CERN, a Forward Physics Facility at the LHC.  At the Energy Frontier, an $e^+/e^-$ Higgs Factory (e.g. FCC-ee, C3, etc) will likely be the next major accelerator facility; interest in discovery machines such as O(10 TeV c.m.e.) muon colliders has also gained significant momentum. These machines' open-ended discovery potential includes sensitivity to a range of particle DM scenarios.  More information can be found in the AF report~\cite{AF-Report}.

\paragraph{Community Engagement and Workforce Development for Dark Matter.}

The last two decades have seen an explosion in the number of physicists engaged in DM detection, and the enthusiasm for the topic has been palpable throughout Snowmass 2021. In order to support this enthusiasm, the community needs to enhance its engagement efforts at all levels of society, most notably with the education system, industry partners, and policy makers.
Outreach widens access to information, and the exciting mystery of DM acts as a powerful recruiting tool for HEP. 

Career pipeline and development are crucial to sustaining such an expansion, which in turn requires a renewed focus on the diversity, equity, inclusion, and accessibility of the field. The historical exclusion of marginalized people from high energy physics is fundamentally harmful to the humanity of individuals who are excited and curious about science. Moreover, creating equal opportunity and equality in particle physics is essential to professional success in our field, ensuring breadth of perspectives and a deep talent pool. Developing a broad talent pool sustains both scientific advancements and the democratic principle of publicly supported and engaged activities.

Equal opportunities in HEP go hand in hand with more traditional priorities. 
The search for DM is highly interdisciplinary and is therefore an excellent training ground for our next generation of scientists. DM searches that rely on quantum sensing and AI/ML align with the need for training in these areas both in HEP and beyond. The long gaps between design and commissioning of large-scale HEP projects such as future colliders can lead to leaks in the pipeline of key knowledge holders. Small- and medium-scale DM projects can bridge these gaps, training scientists across all experimental phases, from design to construction to commissioning and analysis. Thus, beyond their scientific merits, these projects sustain expertise in hardware development, construction, and their interface with science delivery, benefiting the field overall.  The Community Engagement Frontier report~\cite{CEF-Report} addresses these and other workforce questions.

\section{Case Studies}
\label{sec:CaseStudies}

\begin{figure}[htp]
\begin{center}
\includegraphics[width=0.75\textwidth]{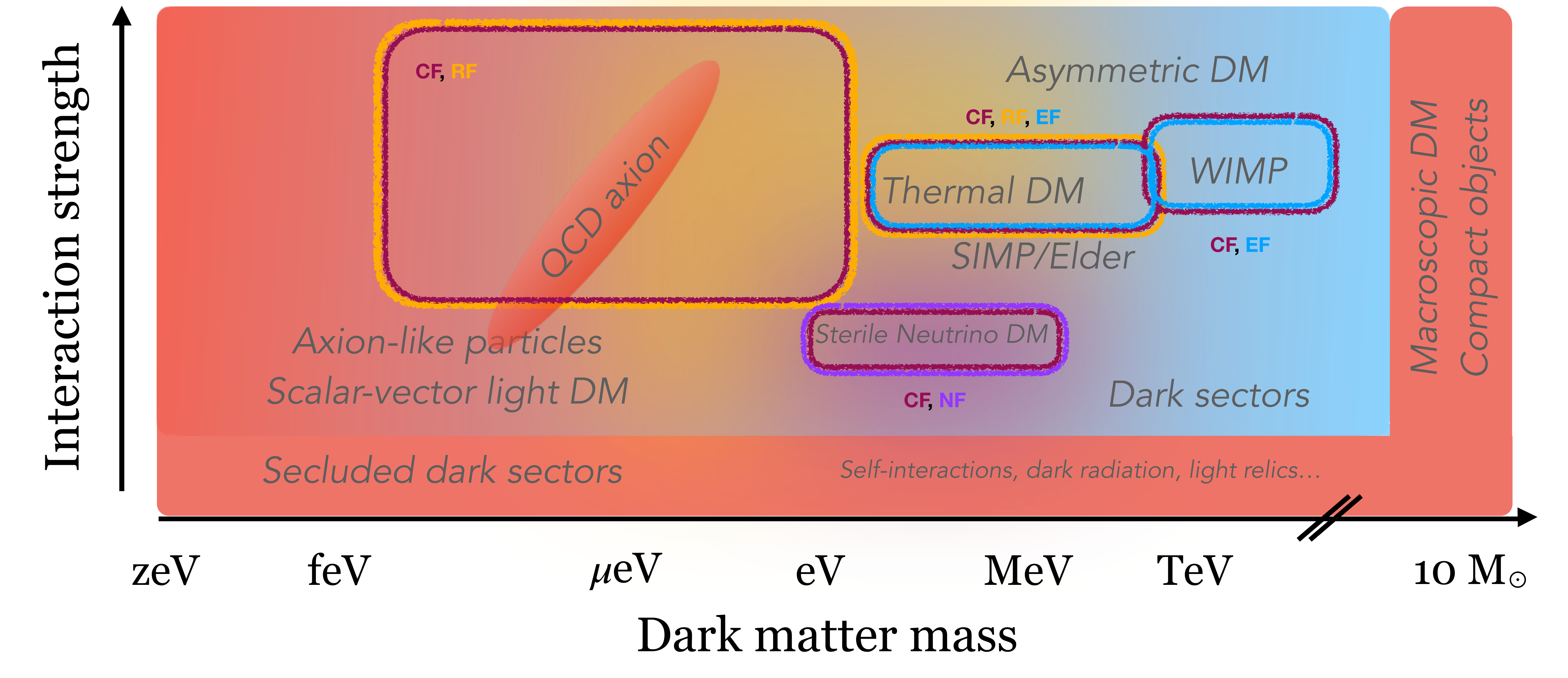}
\vspace{-2em}
\end{center}
    \caption{Summary of case studies presented in this document, shown in the context of a sketch of the coupling-mass plane including the parameter space typical of some of the rich variety of DM theories possible. The shaded colors in this sketch are suggestive of the Frontiers with experiments represented in the case studies in a given region, with color coding specified near the rounded rectangles.}
    \label{fig:allcasestudies}
\end{figure}

Fig.~\ref{fig:allcasestudies} provides a graphical summary of the breadth of theoretical scenarios that can provide DM candidates. The  possibilities span enormous ranges in DM mass and interaction strength.

In the case studies below, we briefly discuss several scenarios to illustrate how complementarity between DM searches could enable discovery of the fundamental nature of DM and allow triangulation of its properties. More detailed versions of these scenarios may be found in Ref.~\cite{Boveia:2022syt}. Additional relevant case studies may be found in Refs.~\cite{BRNreport,Drlica-Wagner:2022lbd}.

\paragraph{Minimal WIMP Dark Matter.}

One of the simplest possibilities for DM involves a new particle multiplet that interacts with SM particles via the weak interaction \cite{Cirelli:2005uq}.  Accidental or imposed symmetries lead to the stability of the lightest particle of this multiplet, providing a suitable DM candidate. The two most-widely studied scenarios are the cases where the DM is part of a Dirac fermion doublet (called the Higgsino) or of a Majorana fermion triplet (Wino); these multiplets appear in supersymmetric (SUSY) theories as superpartners of the SM Higgs/gauge bosons, and so the cases where the DM is close to ``pure Higgsino'' or ``pure Wino'' can be realized in specific regions of the broader  parameter space of SUSY theories.\footnote{Setting aside the connection to supersymmetry, which is not essential, many of the same general principles also apply to larger electroweak multiplets; for a discussion of other options see e.g. \cite{Bottaro:2021snn, Bottaro:2022one}.} Such scenarios are very predictive, with the only free parameter being the DM mass, and this can also be fixed (at 1.1 TeV for the Higgsino and 2.8 TeV for the Wino) by matching the relic density under the assumption of a standard cosmology with thermal freezeout of the DM \cite{Bottaro:2021snn, Bottaro:2022one}.

Indirect searches for gamma-rays and antiprotons already constrain the thermal Wino (e.g.~\cite{Rinchiuso:2018ajn,Cuoco:2017iax}), albeit with significant systematic uncertainties associated with the DM density and cosmic-ray propagation. Future cosmic probes and complementary astrophysical measurements could reduce these uncertainties and strengthen the bounds. In direct detection, the cross section is very small~\cite{Chen:2019gtm}, but large future direct detection experiments such as DARWIN can probe the Wino scenario, with cross-sections still above the neutrino fog \cite{Chen:2018uqz}. The direct detection cross section for the thermal Higgsino lies in the neutrino fog, but CTA should have sensitivity to the indirect-detection signal \cite{Hryczuk:2019nql}. Current colliders can only probe lighter non-thermal Wino- and Higgsino-like particles, but a future multi-TeV lepton or hadron collider could meet the thermal target for both Wino and Higgsino (and eventually even for larger multiplets)~\cite{Bottaro:2021snn, Bottaro:2022one}. A direct or indirect detection during the planning phase of such a collider would provide crucial input to the design. A collider discovery would provide in-depth information on the WIMP’s interactions with SM particles and its associated particle spectra. Alternatively, null results at collider experiments could significantly constrain the interpretation of a putative DM signal from direct or indirect detection. 

\begin{figure}[htp]
\centering
\subfloat[]{\label{subfig:WIMP_ID_Colliders}\includegraphics[width=0.5\textwidth]{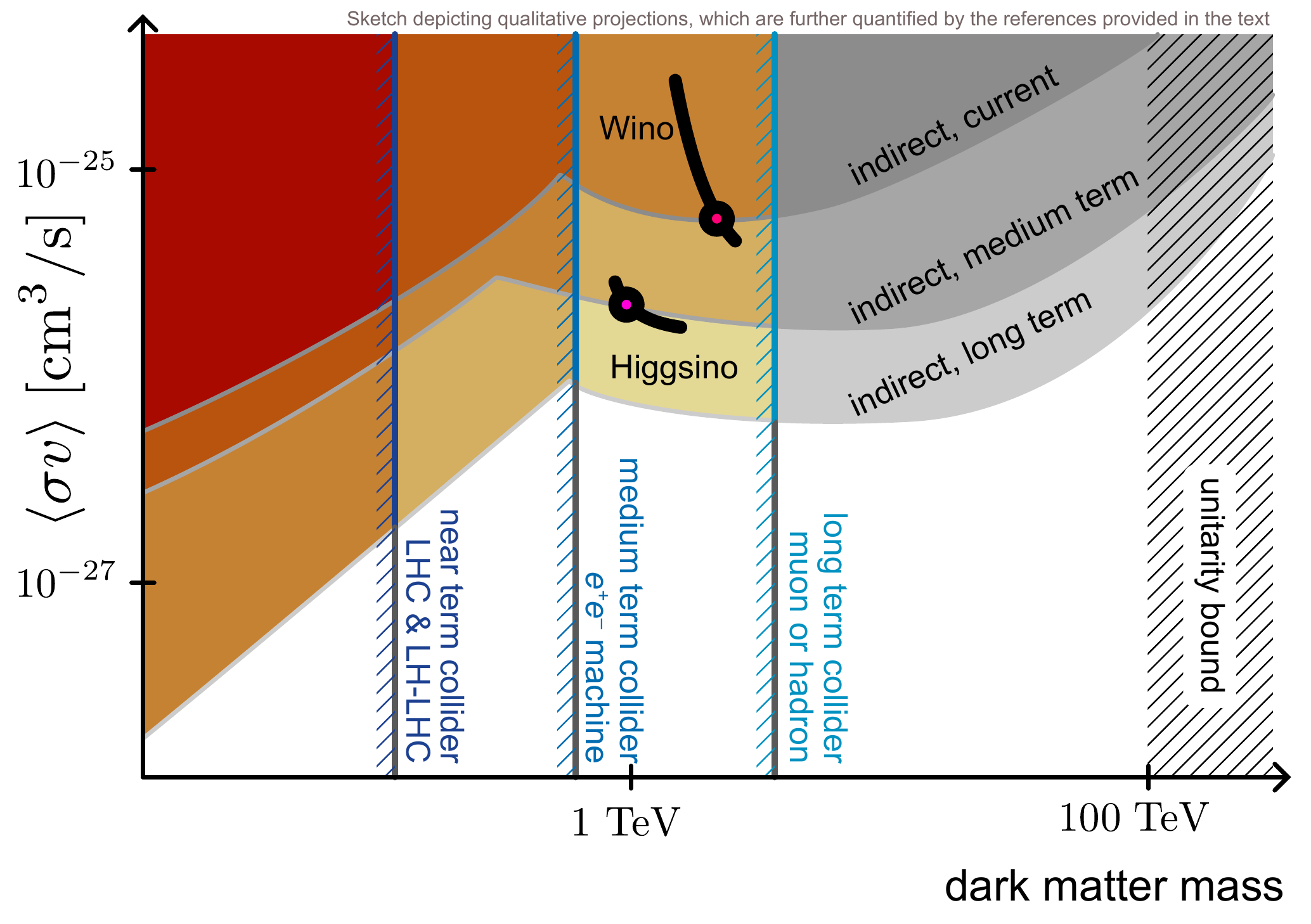}}
\subfloat[]{\label{subfig:BSMMediation}\includegraphics[width=0.5\textwidth]{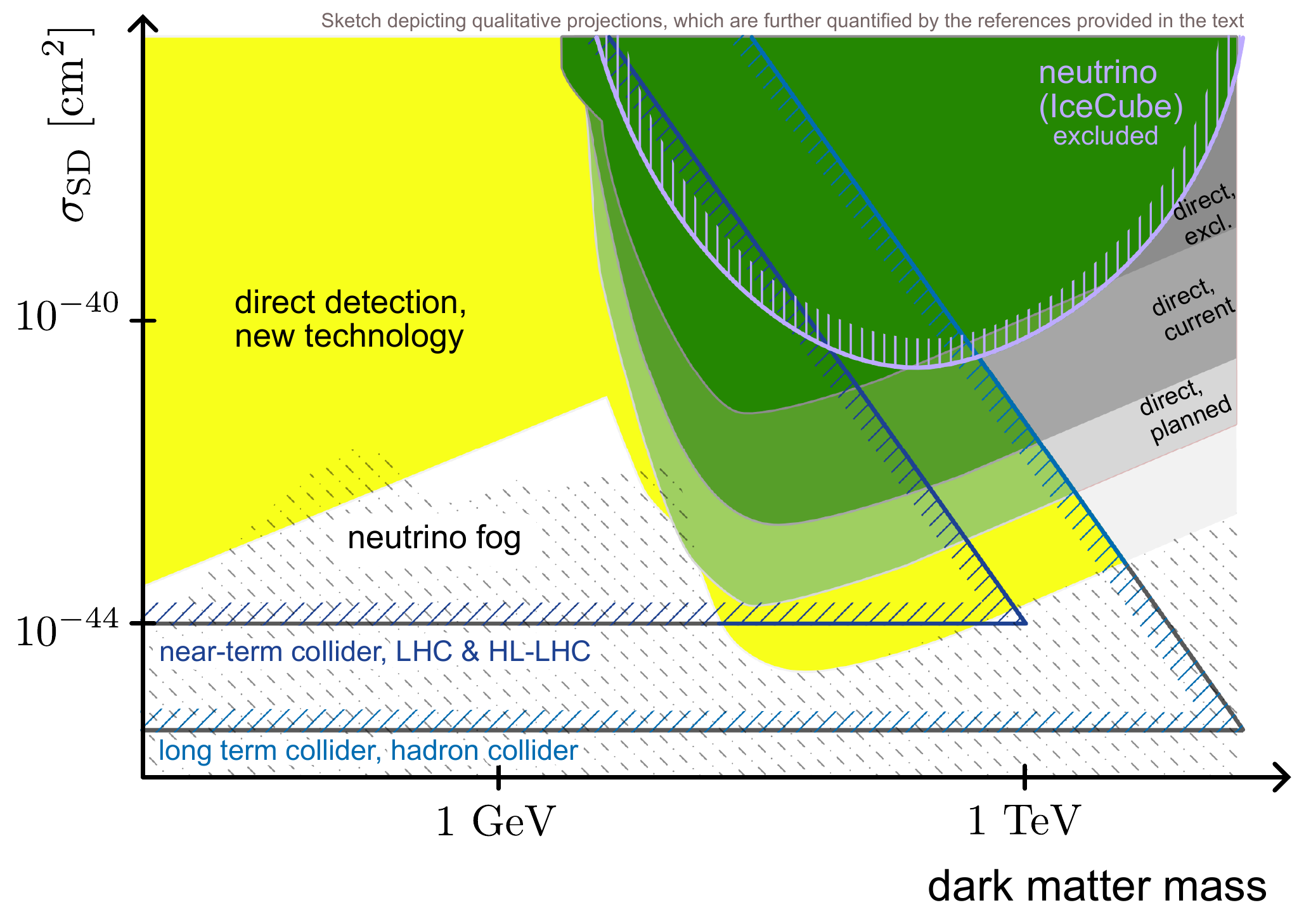}}
\caption{Indicative sketches depicting qualitatively how future collider, indirect detection, and neutrino experiments \protect\subref{subfig:WIMP_ID_Colliders} or collider and direct detection experiments \protect\subref{subfig:BSMMediation} may complement each other during discoveries of Wino/Higgsino DM or of BSM-mediated DM, respectively. The $y$-axis indicates the annihilation cross section to example SM final states in (a) and the spin-dependent scattering cross section on SM targets in (b). Regions where “excluded” is mentioned in the figure have been covered by published results, while other areas depict approximate regions of sensitivity for current and future experiments. 
Regions of overlapping coverage, where complementary observations in both types of experiments would be possible, are indicated by saturated colors. Regions accessible by only one of the two types of experiments are shown in muted colors or grayscale. 
}
    \label{fig:WIMP}
\end{figure}

\paragraph{Generic Beyond the Standard Model (BSM)-mediated and Vector Portal Dark Matter.}
If DM particles are discovered by either Cosmic Frontier or accelerator-based (EF and/or RPF) experiments,  the other technique will be essential to understanding its properties. 
In the case of a CF detection of DM particles, different types of cosmic probes and target materials can shed some light on the nature of DM interactions. However, producing the same kind of DM in the lab, with a known initial state, opens new windows to characterising DM interactions and resolving the roles of related, cosmologically unstable, particles.  Likewise, while a signal of invisible particle production in a fixed target or collider experiment can be related to DM models, a simultaneous discovery in Cosmic Frontier experiments is needed to ascertain the cosmological nature of the DM candidate. 

\begin{figure}[htp]
\centering
\subfloat[]{\label{subfig:DarkPhoton}\includegraphics[width=0.5\textwidth]{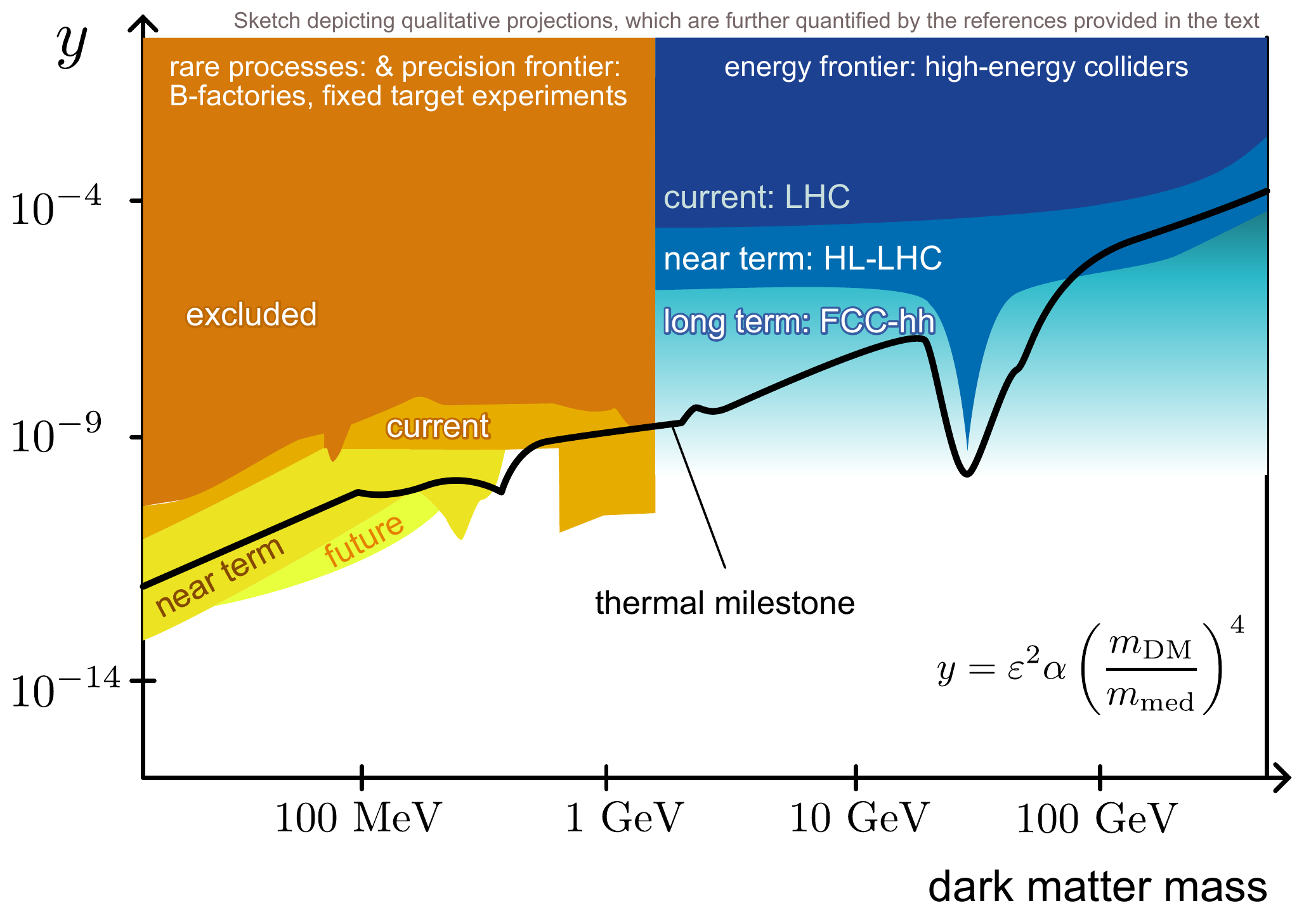}}
\subfloat[]{\label{subfig:SterileNeutrino}\includegraphics[width=0.5\textwidth]{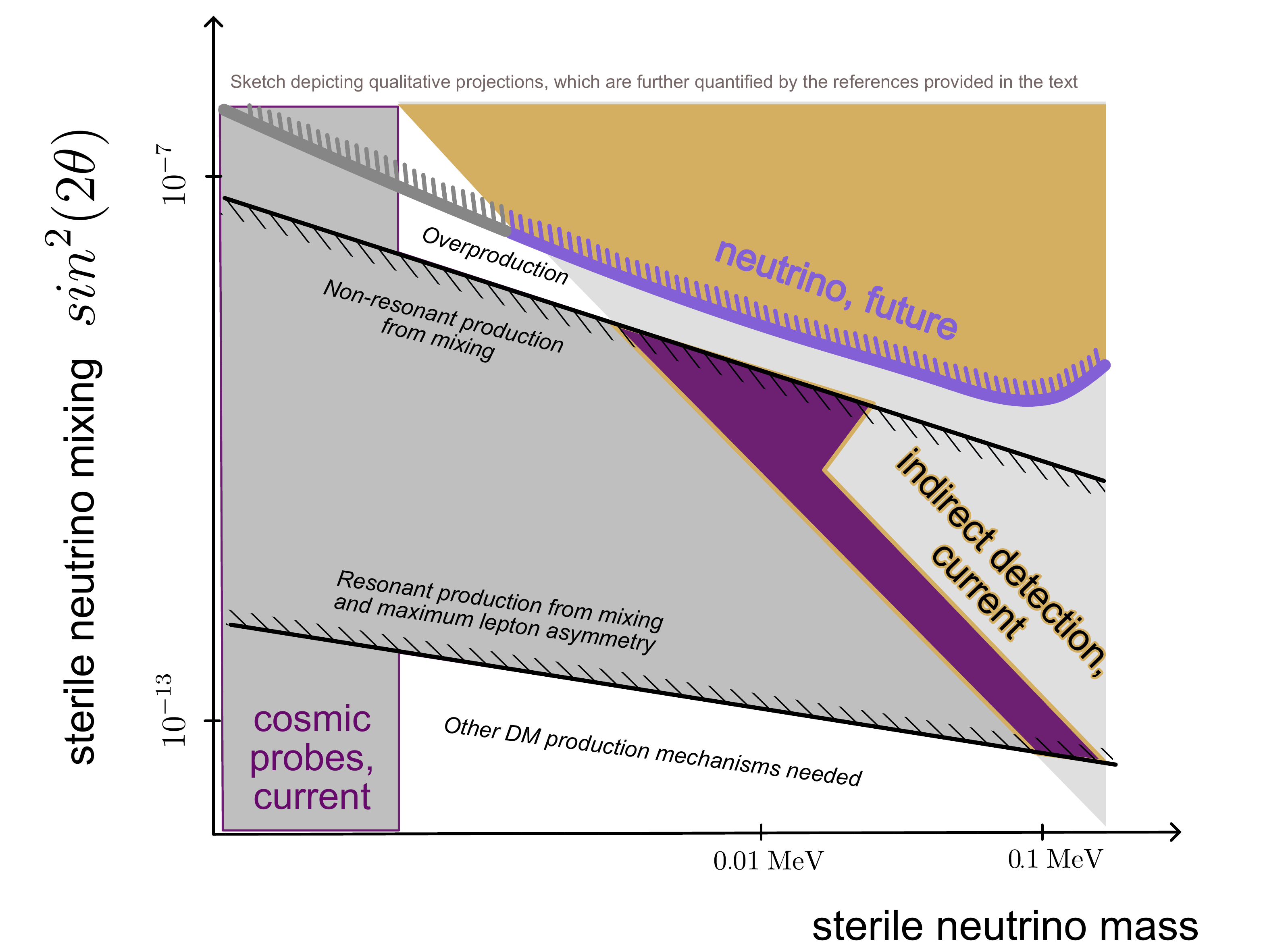}}
\caption{\protect\subref{subfig:DarkPhoton} Sketch of how collider and accelerator experiments together can reach sensitivity across many orders of magnitude of DM mass to couplings expected for thermal-relic vector portal inelastic Dirac DM production (note: lepton colliders not shown). $y$ parameterizes the DM-SM interaction strength and depends on the kinetic mixing parameter $\epsilon$, dark-sector coupling $\alpha$, and masses of the DM ($m_\mathrm{DM}$) and mediator ($m_\mathrm{med}$). 
\protect\subref{subfig:SterileNeutrino} Sketch of constraints on the mass and mixing angle $\theta$ of resonantly produced sterile neutrino DM from indirect detection in X-rays, cosmic probes of small scale structure, and projected sensitivity of tritium beta-decay neutrino experiments.}
    \label{fig:DarkPhotonSterileNeutrino}
\end{figure}

This kind of complementarity is illustrated by theoretical scenarios that extend the WIMP paradigm to include an additional particle beyond the SM that mediates interactions between DM and SM. These mediator particles can decay into both DM and SM particles; searches for each decay mode offer further insight into the DM-SM interaction.  
A thermal history for the DM candidates in the early universe can be attained depending on the coupling types and strengths of the mediator (or \textit{portal}) particle, as well as on the mediator and DM particle masses. 
Specific realizations of these models are used as benchmarks in e.g.~Refs.~\cite{Abercrombie:2015wmb,Energy-Frontier-Report,Beacham:2019nyx,Gori:2022vri}. 
Figures \ref{fig:WIMP} \protect\subref{subfig:BSMMediation} and  \ref{fig:DarkPhotonSterileNeutrino} \protect\subref{subfig:DarkPhoton} illustrate complementarity across CF, EF and RPF experiments in terms of both opportunities for simultaneous discovery (CF1/EF10) and complementary discovery sensitivity (EF10/RF6); we also refer the reader to Fig.~1-1 and Case studies 1 and 2 of~\cite{BRNreport} for discussion of the CF1/RF6 complementarity for low-mass DM. Further insights into these and related portal DM models can be gained from MeV-energy indirect detection (CF1) and probes of DM self-interaction (CF3).

\paragraph{Sterile Neutrino Dark Matter.}

Figure \ref{fig:DarkPhotonSterileNeutrino} \protect\subref{subfig:SterileNeutrino} is an illustration of the complementarity between cosmological, astrophysical, and laboratory searches for sterile neutrino DM (as compiled by \cite{Abazajian:2022ofy}), with complementary reach provided by a combination of indirect DM searches (CF1; \cite{Cooley:2022ufh}), cosmic probes of structure formation (CF3; \cite{Drlica-Wagner:2022lbd}), and laboratory neutrino facilities (NF3; \cite{Coloma:2022dng}). 
Indirect detection experiments searches for X-ray lines originating from the decay of keV mass sterile neutrinos. 
Current constraints come from Chandra, XMM-Newton, NuSTAR, INTEGRAL, and Fermi GBM observations of the various astrophysical systems (e.g., the Milky Way, M31, dwarf galaxies, and galaxy clusters) \cite{Horiuchi:2013noa,Boyarsky:2005us,Boyarsky:2006zi,Abazajian:2006jc,Malyshev:2014xqa,Sicilian:2020glg,Foster:2021ngm,Roach:2019ctw}. 
Future X-ray facilities such as {\it XRISM}, {\it Athena}, and the WFM instrument aboard the {\it eXTP} X-ray Telescope could increase sensitivity to mixing angle by orders of magnitude \cite{Neronov:2015kca,Zhong:2020wre,Malyshev:2020hcc,Ando:2021fhj}.
Cosmological constraints are due to the suppression of DM structure that occurs for producing keV-mass sterile neutrinos with a ``warm'' initial momentum distribution \cite{Bullock:2017xww,Schneider:2016uqi}.
These constraints will improve by orders of magnitude as measurements of the least massive DM halos improves with DESI \cite{Valluri:2022nrh}, Rubin LSST \cite{Mao:2022fyx}, and future cosmological survey experiments \cite{Chakrabarti:2022cbu}.
Laboratory searches (e.g., Katrin, TRISTAN, BeEST, and HUNTER) \cite{Merle:2017jfn,KATRIN:2018oow,Friedrich:2020nze,Martoff:2021vxp} have different dependencies on the model behavior of sterile neutrinos in the early Universe (e.g., in cosmologies with large lepton asymmetry, low reheating temperature and/or neutrino non-standard interactions) making them highly complementary to indirect and cosmological searches \cite{Gelmini:2019clw, Gelmini:2019wfp}. Below the lowest solid line, achieving the correct relic density of sterile neutrino DM requires other production mechanisms beyond active-sterile mixing; these generally involve heavy BSM particles, which can be probed at the Energy Frontier. 

\paragraph{Wave-like Dark Matter: QCD Axion Discovery.}
The discovery of a QCD axion by a direct detection experiment such as DMRadio-$m^3$ with a mass $< 1\,\mu$eV can be used to illustrate of the wide-ranging implication of a wave-like DM discovery. The detailed spectral measurements from direct axion searches would almost instantaneously provide a measurement of the velocity distribution of DM in the halo. These measurements and subsequent measurements of the position distribution can then be compared to the results from cosmic probes of DM.  Since axions with masses below $< 1\,\mu$eV imply additional fields at the time of inflation, CMB B-modes would then be out of range for next generation CMB experiments. A possible discrepancy between such measurements would open the door to significant changes in our understanding of particle physics and cosmology.

Most direct detection experiments use the axions' coupling to photons. A precision measurement of this coupling, or the coupling to other parts of the SM, can disentangle which category of QCD axion or axion-like particle has been discovered. There is complementarity here with other table-top precision measurements. The discovery of a QCD axion implies additional particles, e.g.~KSVZ models predict additional quarks and DFSZ predict an expanded Higgs sector. Either would be strong motivation for higher-energy colliders and would guide the design of such efforts.

\section{Conclusions}

DM presents a fundamental puzzle to particle physics. The space of viable, theoretically-motivated candidates is enormous and multi-dimensional, spanning many orders of magnitude in mass and interaction strength. To make progress on this challenging problem, maximize the chances of a transformative discovery, and fully elucidate the properties of DM and related new physics in the event of such a discovery, we advocate for a cross-Frontier effort incorporating multiple complementary approaches to the problem. The suite of approaches discussed in this summary, and in the various topical group and Frontier reports, will allow us to delve deep into highly compelling, long-standing, and well-studied scenarios for the nature of DM, and simultaneously to open up our search to a wide and less-explored space of exciting and well-motivated possibilities. A decade of coherent cross-Frontier DM exploration is an opportunity that should not be missed.

\printbibliography

\section{Acknowledgments}
The authors would like to thank the following people for their input in the discussions and cross-frontier meetings that led to this report, including inputs to the figures: Suchita Kulkarni, Ben Loer, Bjoern Penning, Hai-Bo Yu. 


\end{document}